\documentclass{ws-procs11x85}
\pdfpagewidth=\paperwidth
\pdfpageheight=\paperheight
\usepackage{ws-procs-thm}
\usepackage[utf8]{inputenc}
\usepackage[T1]{fontenc}
\usepackage{graphicx}
\usepackage{microtype}

\graphicspath{{figures/}}

\let\CLARAoriginalsubsection\subsection
\renewcommand{\subsection}[1]{%
  \par\addvspace{7pt}%
  \CLARAoriginalsubsection{#1}%
}

\begin{document}


\raggedbottom
\setlength{\abovedisplayskip}{9pt plus 2pt minus 1pt}
\setlength{\belowdisplayskip}{9pt plus 2pt minus 1pt}
\setlength{\abovedisplayshortskip}{7pt plus 1pt minus 1pt}
\setlength{\belowdisplayshortskip}{7pt plus 1pt minus 1pt}
\setlength{\jot}{3pt}

\title{CLARA: Clarification of Language Ambiguity through Result Analysis for Natural-Language Cancer Genomics Queries}

\author{Pratyush Kumar Shukla$^\dag$, Manveer Singh Tib and Siddhant Garg}

\address{Independent Researchers\\
$^\dag$E-mail: pratyushs2009@gmail.com}

\begin{abstract}
A natural language interface can be used to make cancer genomics databases easier to use, but even if a question is perfectly fluent, its scientific meaning can be ambiguous. We propose CLARA, a framework that represents a question as a typed scientific query specification, considers a few possible interpretations, executes them, and asks for clarification when the estimates diverge. CLARA was assessed on mutation-prevalence contrasts among eight TCGA PanCancer Atlas cohorts and a 30-gene panel. This benchmark consisted of 330 unique executable contrasts varying in mutation scope, assay denominator, and sample context; 115 contrasts were result-sensitive and 215 were result-stable, per the preregistered definition of relative divergence greater than 0.10 or absolute divergence greater than 5 percentage points. An independently implemented pandas execution engine perfectly replicated all 660 results from the SQLite engine. In a separate 120 question LLM-generated, manually vetted language stress test, CLARA recognized all 60 result-sensitive contrasts and needlessly clarified 13 of 60 stable contrasts (accuracy 89.2\%, sensitivity/recall 100\%, specificity 78.3\%). Standalone machine learning had superior overall accuracy (97.5\%) but missed one critical contrast. This demonstrates that downstream execution can distinguish consequential from inconsequential ambiguity and reveal an explicit trade-off between safety and burden. 
\end{abstract}


\copyrightinfo{\copyright\ 2026 The Authors. Open Access chapter published by World Scientific Publishing Company and distributed under the terms of the Creative Commons Attribution Non-Commercial (CC BY-NC) 4.0 License.}

\section{INTRODUCTION}

Even a simple question such as “How common are TP53 mutations in lung adenocarcinoma?” leaves several decisions implicit. The analysis may include patients or samples, define mutation broadly or restrict to truncating mutations, include only cases profiled for mutation or treat the full cohort as the denominator, and include primary, recurrent or metastatic specimens. Each interpretation can be compiled into valid SQL. So the danger is not just in syntactically incorrect code. A system can produce a precise, reproducible number that answers the wrong scientific question.

Larger text-to-SQL benchmarking efforts have pushed the field of cross-domain semantic parsing further, which includes the works of Spider \cite{ref1}, its contextual and conversational extensions SParC \cite{ref2} and CoSQL \cite{ref3}, and more recent enterprise-based benchmarks like Spider 2.0 \cite{ref4}. The biomedical domain has seen the use of natural language database querying for applications in electronic health records \cite{ref5}, clinical cohort definition \cite{ref6,ref7}, and knowledge bases for question answering such as PubMedQA \cite{ref8} and BioASQ \cite{ref9}. However, in most benchmarking efforts, the assumption is that there is only one executable target for each question. 

NaLIR performed translation of complex questions via limited interaction \cite{ref10} while parser-independent interaction was used for asking users targeted multiple choice questions \cite{ref11}. There is recent research that identifies ambiguous and unrecognizable text-to-SQL requests \cite{ref12}. Execution has been used to filter invalid programs \cite{ref13}, constraints to produce only valid SQL \cite{ref14}, semantic equivalence testing by executing test suite \cite{ref15}, and detecting potential parser errors \cite{ref16}. 

The patient-sample-molecular profile hierarchy means that the denominator is part of the scientific definition, not just its implementation. Records that are missing cannot be assumed to be negative observations in any database with incomplete information \cite{ref17}. There are resources for cancer genomics like cBioPortal that make multiple genomic profiles available \cite{ref18,ref19}; there are resources like OncoTree that provide standard terminologies for diseases \cite{ref20} and resources like TCGA that provide harmonized data for pan-cancer clinical information \cite{ref21}. 

Thus, we introduce CLARA - Clarification of Language Ambiguity through Result Analysis. CLARA transforms the user's question into a typed scientific query specification, picks an open-ended semantic dimension among them, generates the set of plausible candidate specifications, evaluates them all, and compares their outcomes. If the alternatives differ in a material way, CLARA asks for clarification; otherwise, CLARA gives the answer, leaving the user alone but having an explicit specification of their intended interpretation.

First, we present an executable specification of cancer mutation-prevalence question with regard to analysis unit, assay eligibility, sample context and mutation type. Second, we formulate the notion of clarification as a result-aware decision instead of a simple linguistic one. Third, we create a multicohort benchmark of 330 unique executable contrasts with verification of all candidate results. Finally, we assess safety-efficiency tradeoff in the framework of controlled confirmatory benchmark and language stress test separately. 

\section{METHODS}

\subsection{System overview}

CLARA analyzes a natural-language prevalence question by extracting entities, constructing typed specifications, identifying ambiguity, generating and executing candidate specifications, and determining whether to answer or request clarification. Its modular design separates language-based detection of the cohort, gene, and semantic mechanism from deterministic candidate evaluation.

\begin{figure}[!tbp]
\centering
\includegraphics[width=0.78\textwidth]{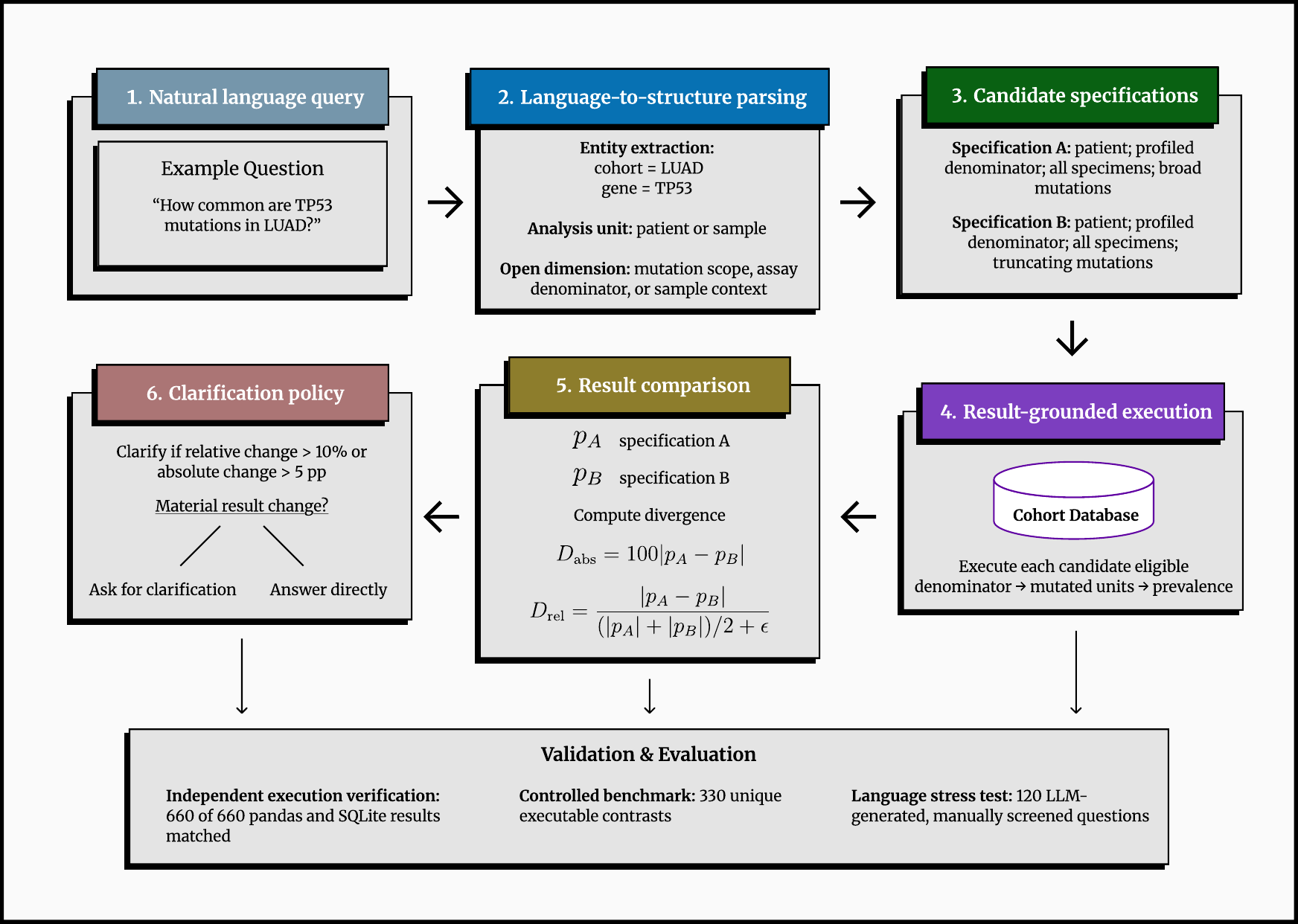}
\caption{CLARA Workflow. A question about cancer genomics is formulated using natural language and translated into a typed specification.}
\label{fig:clara-pipeline}
\end{figure}
\subsection{Data sources and cohort construction}

We used eight TCGA PanCancer Atlas studies distributed through cBioPortal: bladder urothelial carcinoma (BLCA), breast invasive carcinoma (BRCA), colorectal adenocarcinoma (COADREAD), head and neck squamous cell carcinoma (HNSC), lung adenocarcinoma (LUAD), prostate adenocarcinoma (PRAD), skin cutaneous melanoma (SKCM), and uterine corpus endometrial carcinoma (UCEC). Labels for cohorts were normalized to study identifiers and short names. The consolidated database included 411, 1,084, 594, 523, 566, 494, 448, and 529 samples, respectively; SKCM had 442 unique patients, as multiple patients contributed repeated specimens. We also explicitly retained mutation profiling and primary-sample indicators.

A fixed 30-gene pan-cancer panel was used: TP53, KRAS, PIK3CA, PTEN, APC, BRAF, NRAS, NF1, EGFR, ARID1A, KDM6A, NOTCH1, RB1, CDKN2A, FBXW7, CTNNB1, ATM, BRCA1, BRCA2, SMAD4, STK11, KEAP1, ERBB2, FGFR3, GATA3, SPOP, FOXA1, IDH1, VHL, SETD2. To prevent selection of genes based on the observed result, all eight cohorts were subjected to the same panel. Two nested definitions were mapped to mutation records. The broad definition encompassed protein-altering coding events and excluded silent and noncoding entries. The truncating definition included nonsense, frameshift, splice-site, translation-start and nonstop events.

\begin{table}[!tbp]
\tbl{Multicohort data and executable benchmark composition.}
{\begin{tabular}{@{}lrrr@{}}
\toprule
Ambiguity method & Total & Result-sensitive & Result-stable\\ \colrule
Mutation scope & 82 & 79 & 3\\
Assay denominator & 124 & 17 & 107\\
Sample context & 124 & 19 & 105\\ \botrule
\end{tabular}}
\label{tab:cohort-composition}
\end{table}

\subsection{Typed scientific query specification}
Six fields of the immutable ValidationSpec were an executable interpretation:
\begin{itemlist}
\setlength{\itemsep}{0pt}
\setlength{\parsep}{0pt}
\item study\_id: the cancer cohort;
\item gene: the queried gene;
\item analysis\_unit: patient or sample;
\item assay\_requirement: whether the denominator is restricted to molecularly profiled units;
\item sample\_context: primary tumors only or all available tumor specimens; and
\item mutation\_scope: broad protein-altering or truncating mutations.
\end{itemlist}

The execution engine constructed an eligible denominator before counting mutation-positive units. For patient-level analysis, both numerator and denominator used distinct patient identifiers; for sample-level analysis, both used distinct sample identifiers. When assay\_requirement was true, only profiled samples were eligible. When sample\_context was primary, only primary-tumor samples were eligible. The numerator was then intersected with the denominator, preventing an ineligible mutation record from entering the count. Prevalence was computed as n\_mutated/n\_denominator. A unit lacking a mutation record was considered negative only when it belonged to the eligible profiled denominator; unprofiled units were not silently treated as wild type.

\subsection{Ambiguity mechanisms and candidate generation}

Scope of mutation was at a patient level, with mutation profiling, all specimen types, and broad protein-altering versus truncating mutations.

The assay denominator mechanism was to perform the analysis at the patient level, to use broad protein-altering mutations and all specimen types, and to compare a profiled-only denominator to the entire cohort denominator.

Sample context kept the analysis at the sample level, required mutation profiling, used broad protein-altering mutations, and only compared primary tumors with all tumor samples.

We retained cohort-gene pairs if they satisfied the structural eligibility criteria specific to the mechanism and both candidate specifications were executable. The final registry included 330 unique scientific contrasts, including 82 mutation-scope, 124 assay-denominator, and 124 sample-context contrasts. The primary statistical unit was the individual executable contrast, with paraphrases as grouped language realizations and not as independent biological observations.

\subsection{Result-divergence criterion}

Let $p_A$ and $p_B$ be the prevalences generated by the two candidate specifications. We calculated absolute divergence, in percentage points,

\begin{equation}
\begin{aligned}
D_{\mathrm{abs}} &= 100 \times \lvert p_A - p_B \rvert,\\
D_{\mathrm{rel}} &= \frac{\lvert p_A - p_B \rvert}{((\lvert p_A \rvert + \lvert p_B \rvert)/2) + \epsilon},
\end{aligned}
\label{eq:divergence}
\end{equation}

where $\epsilon = 10^{-12}$ to prevent division by zero. We flagged a contrast as result sensitive if D\_rel > 0.10 OR D\_abs > 5 percentage points. If not, it was result-stable. The OR rule was selected to guarantee that relative divergence could detect a small absolute change at low prevalence and that a clinically evident absolute shift would not be masked by a high baseline prevalence. Threshold-robustness analyses changed both elements after primary analysis but did not affect the registered labels.

\subsection{Natural-language translation and frozen hybrid policy}

Boundary-aware alias matching was used to extract cohorts and gene entities. Boundary constraints were necessary because the cohort abbreviation BRCA appears in the gene symbols BRCA1 and BRCA2. A deterministic rule parser looked for exactly one unresolved dimension while checking that the other specification fields were explicit. A statistical parser used a scikit-learn pipeline of word TF-IDF features over 1-3-grams, character-boundary TF-IDF features over 3-5-grams, and class-balanced multinomial logistic regression with a fixed random seed and a maximum of 3,000 iterations. The auxiliary training set had 4,320 generated questions, 1,440 per mechanism, with no exact overlap with the evaluation questions.

The final hybrid policy was frozen post-development diagnostics. If the rule parser found exactly one mechanism, it used the output. Otherwise, the statistical parser was used if its maximum class probability was at least 0.60. If neither condition is satisfied, CLARA asks for clarification. Once a mechanism is identified, CLARA executes the corresponding candidate specifications and evaluates the candidate results using the registered sensitivity criterion. 

\subsection{Benchmark design and leakage controls}

There were four templated paraphrases for each of the 330 contrasts, making 1,320 questions in the development benchmark. Initial diagnostics indicated a cohort-alias collision with BRCA and that some mechanisms had been missed by the rule parser. We held boundary matching constant and froze a post-diagnostic change prior to confirmatory testing. We then generated three new confirmatory templates per mechanism, for a total of 990 questions over the same 330 contrasts. These questions did not have any exact overlap with development templates or auxiliary training questions. Variant 1, one question per contrast, was prespecified as the primary controlled confirmatory evaluation. Variants 2 and 3 were kept as clustered paraphrase robustness checks. Since this benchmark was based on diagnostic amendments and applied the same contrast definitions, we speak of the internal post-diagnostic confirmation rather than external validation.

In order to analyze more varied language, we chose 120 unique contrast sets, split evenly into 60 result-sensitive and 60 result-stable examples. Three new LLM sessions created 40 questions each using assignment files that included just the cohort, the gene, the type of ambiguity desired, and the constraints to hold other variables constant. The sessions received no labels of results, prevalence, parsing rules, previous questions, or frozen outputs. The pool included no empty questions, exact repeats, or question pairs above the prespecified duplicate threshold of cosine similarity. Two members of the project reviewed the questions and accepted all 120 unmodified. They did not record categorical form for any items, hence we cannot report inter-rater reliability, and we will use this resource as a language stress test of LLM output.

\subsection{Baselines and endpoints}

Rule used just the deterministic mechanism parser and clarified cautiously any translation that was not valid. Standalone ML used the statistical mechanism parser for each question asked. The mechanism prior clarification policy clarified any question where the scope was mutations and gave an answer without any execution on any other mechanisms. The language-only clarification policy clarified all questions which were ambiguous. The fixed default clarification policy never clarified.

The first criterion for safety was the critical error rate, defined as the percentage of result-sensitive queries that were answered directly without clarification. Unnecessary clarification was the criterion for efficiency, defined as the percentage of result-stable queries that were flagged for clarification. Other metrics were accuracy, precision, recall, specificity, F1, clarification rate, and immediate answer rate.

\subsection{Statistical analysis and reproducibility}

Two-tailed 95\% Clopper–Pearson exact confidence intervals were generated for binary proportions. Systems were compared using paired results of exact McNemar tests; secondary comparisons were corrected using Benjamini–Hochberg adjustment. For the controlled benchmark, 10,000 cluster-bootstraps for cohort–gene families were conducted. For the 120-question stress test, where each question constituted a different contrast selected, 10,000 non-parametric bootstraps were conducted at the question level. A perfect classifier gives a degenerate bootstrap confidence interval; hence, emphasis is placed on exact binomial intervals.

The SQL execution engine was separately re-implemented in pandas after the results were frozen. This separate implementation reconstructed eligible, mutation positive, numerator, denominator, and prevalence, respectively, without reusing the SQL query. This was done for all of the 330 contrasts with respect to two candidate executions. SHA-256 hash values were computed for protocol files, benchmark files, predictions, figures, and statistical results. Code and supplementary materials are available from the corresponding author upon request.

\section{RESULTS}

\subsection{Many linguistic ambiguities were result-stable}

For the eight cohorts, there were 330 distinct executable comparisons in the benchmark, 115 (34.8\%) of which were result-sensitive and 215 (65.2\%) of which were result-stable based on the registered criterion. Sensitivity greatly differed depending on mechanism. Mutation scope was sensitive in 79 out of 82 comparisons (96.3\%), with a median relative divergence of 0.848 and median absolute divergence of 5.19 percentage points. Assay denominator was sensitive in 17 out of 124 comparisons (13.7\%); it had a median relative divergence of 0.015 and median absolute divergence of 0.083 percentage points. Sample context was sensitive in 19 out of 124 comparisons (15.3\%), with a median divergence of 0.00 because few non-primary samples existed compared to the number of primary tumors within most cohorts.

\begin{figure}[!tbp]
\centering
\includegraphics[width=0.62\textwidth]{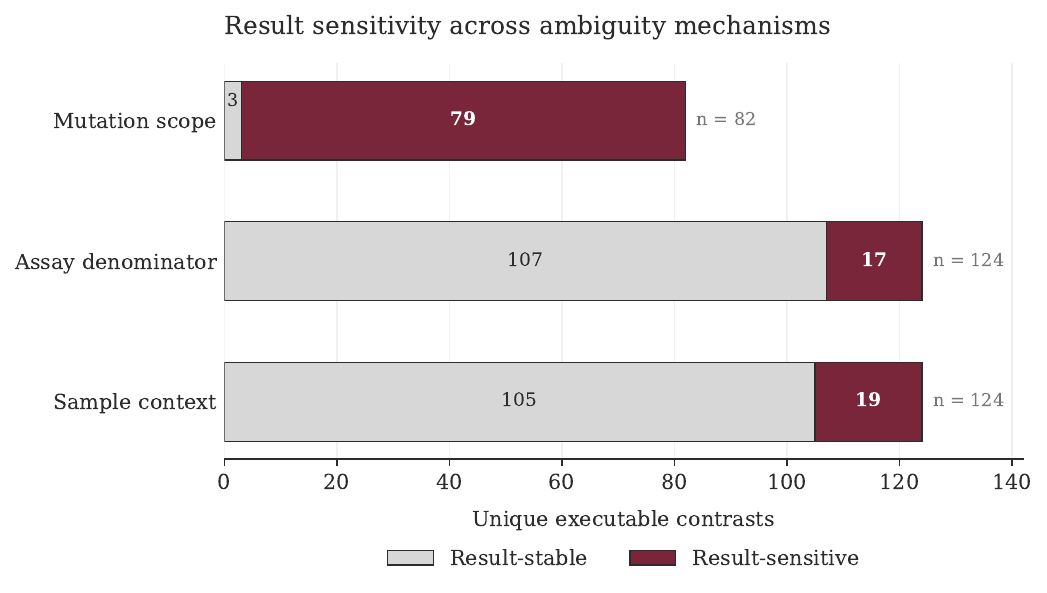}
\caption{Sensitivity of results under different types of ambiguity mechanisms.}
\label{fig:benchmark-composition}
\end{figure}
The former would mean interrupting the user in all 215 stable distinctions, and the latter would mean providing an answer based on an unconfirmed default for all 115 sensitive distinctions. 

\subsection{Independent execution verification reproduced every candidate}

The independent pandas engine was run on 660 candidate specifications, two per contrast. It agreed with the SQLite engine on numerator count, denominator count, and prevalence for all 660 trials, without any disagreements. Sensitivity was measured for a range of tolerance values for relative and absolute divergence. As anticipated, the sensitivity of the contrasts was less when tolerance was higher, but the point of transition was in an area where the mutation-scope contrasts tended to be sensitive and the others to be stable. 

\subsection{Controlled confirmatory evaluation validated pipeline consistency}

The confusion matrix had 215 true negatives, zero false positives, zero false negatives, and 115 true positives. Machine learning alone was able to correctly classify all 330 contrasts too. This means that the controlled benchmark proves internal consistency of the frozen translation-execution pipeline but does not prove any superior accuracy to machine learning or language generalization.

CLARA accuracy had a 95\% exact confidence interval of 98.9–100.0\%; the 95\% confidence intervals for recall and specificity were 96.8–100.0\% and 98.3–100.0\%, respectively. Both critical error and unnecessary clarification probabilities were zero, with the upper 95\% exact limit being 3.16\% and 1.70\%, respectively. 

\subsection{Language translation was accurate at the entity level but mechanism resolution remained imperfect}

The distinct language stress test consisted of 120 problems: 27 mutation scope, 45 assay denominator, and 48 sample context problems, evenly split into 60 sensitive and 60 stable contrasts. Boundary-aware cohort extraction and boundary-aware gene extraction were accurate for all 120 out of 120 problems. Accuracy for the mechanism was 40.0\% using rules alone, 93.3\% using ML alone, and 80.8\% using the hybrid mechanism label. The hybrid mechanism used rules for 48 problems, ML for 50, and conservative clarification for 22. 

\subsection{CLARA eliminated observed critical misses but clarified more stable questions}

For the language stress test, CLARA generated 47 true negatives, 13 false positives, 0 false negatives and 60 true positives. The accuracy of the model was 89.2\% (95\% exact CI, 82.2–94.1\%), the recall for sensitive questions was 100.0\% (94.0–100.0\%) and the specificity for stable questions was 78.3\% (65.8–87.9\%). The critical error rate observed was 0\% (upper 95\% exact limit, 5.96\%). The unnecessary clarification rate was 21.7\% (95\% exact CI, 12.1-34.2\%). 

Using standalone ML, there were 58 true negatives, two false positives, one false negative and 59 true positives, with an accuracy of 97.5\%, recall of 98.3\% and specificity of 96.7\%. Paired analysis revealed that standalone ML was correct on 11 questions that CLARA missed and CLARA was correct on one question that standalone ML missed (exact McNemar P = 0.00635 after correction). Hence, CLARA should be understood as a conservative policy and not as a best classifier.

\begin{table}[htbp]
\vspace{6pt}
\tbl{Final policy performance in the 120 question language stress test.}
{\resizebox{0.68\textwidth}{!}{%
\begin{tabular}{@{}lrrrrrrr@{}}
\toprule
System & TN & FP & FN & TP & Accuracy & Critical error & Unnecessary clarification\\ \colrule
CLARA & 47 & 13 & 0 & 60 & 89.2\% & 0.0\% & 21.7\%\\
Standalone ML & 58 & 2 & 1 & 59 & 97.5\% & 1.7\% & 3.3\%\\
Rule only & 29 & 31 & 0 & 60 & 74.2\% & 0.0\% & 51.7\%\\
Mechanism prior & 29 & 31 & 19 & 41 & 58.3\% & 31.7\% & 51.7\%\\
Language-only & 0 & 60 & 0 & 60 & 50.0\% & 0.0\% & 100.0\%\\
Fixed default & 60 & 0 & 60 & 0 & 50.0\% & 100.0\% & 0.0\%\\ \botrule
\end{tabular}}}
\label{tab:language-results}
\end{table}
\vspace{-17pt}

\begin{figure}[!tbp]
\centering
\includegraphics[width=0.62\textwidth]{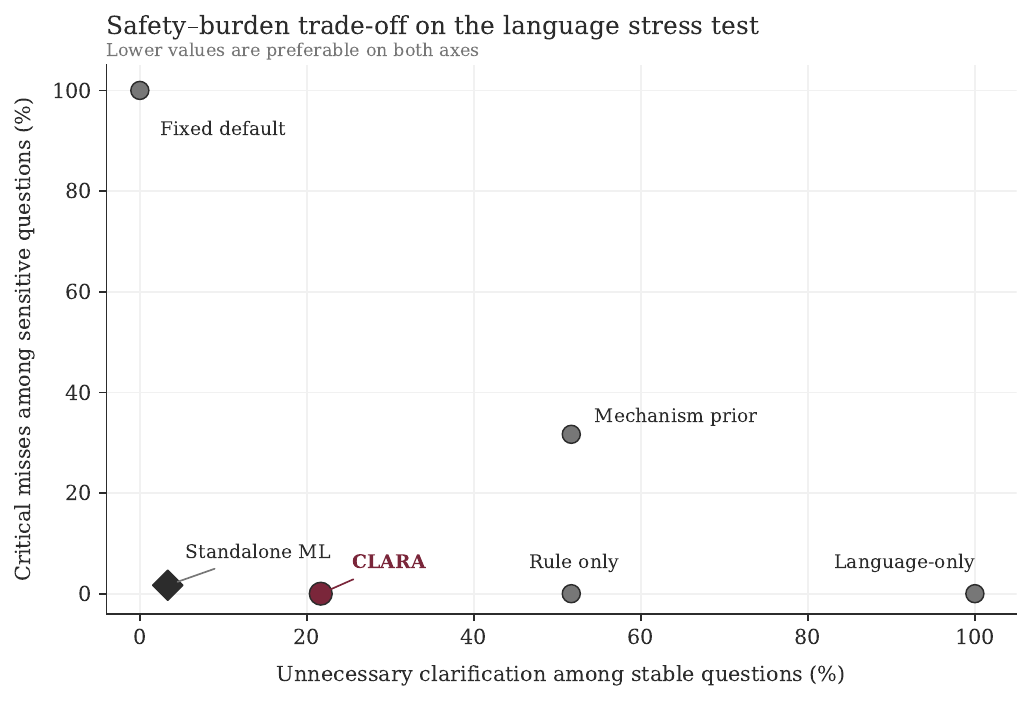}
\caption{Trade-off between safety and burden on the language stress test.}
\label{fig:language-tradeoff}
\end{figure}
\subsection{Errors were concentrated in assay-denominator language}

CLARA had a perfect score on both 27 questions on scope of mutations and 48 questions on context of samples. CLARA made 13 mistakes in 45 questions on assay denominators, and all of those mistakes were unnecessary clarifications. In that particular mechanism, CLARA gave 15 true negatives, 13 false positives, 0 false negatives, and 17 true positives with 71.1\% accuracy and 46.4\% unnecessary clarification rate. 

The nature of the post-evaluation error audit was descriptive. Reviewing these results did not lead to a modification of the parser rules, thresholds, confidence cutoffs, questions and labels.

\section{DISCUSSION}

Our main observation was not that ambiguities are rare but rather that linguistic ambiguity and analytical implication are two distinct variables. Out of 330 executable comparisons, roughly two-thirds were invariant with respect to the accepted tolerance, while one-third was significantly altered. This depended greatly on the scientific mechanism behind the phenomenon being studied. Definitions of broad vs. truncating mutations tended to alter frequencies, while denominator and specimen-based approaches were mostly equivalent, though in this specific case, there could be exceptions with scientific importance. Therefore, a good interface should be able not only to detect but also to evaluate the significance of the ambiguity.

\subsection{Result-aware clarification as a decision layer}

CLARA can be seen as an extra layer that comes between semantic parsing and an answer. Semantic parsing looks for the most probable question, constrained decoding finds the executable question, and error detection evaluates the probability of the generated question being wrong. CLARA poses one more question, which is as follows: amongst the possible valid questions, is the answer stable regarding the ambiguity? The given approach retains the usefulness of current parsers but modifies the interaction strategy. The reliable parser will still use the result analysis technique because the correctness of the wording does not mean that the scientific answer will be stable. On the contrary, an ambiguous question does not necessarily imply interaction.

This model is especially appropriate for those contexts where there is a small number of potential decisions, each having an executable meaning. An instance of such a model would be cancer mutation prevalence because the patient/sample counting, eligibility, sample context, and mutation definition can all be expressed straightforwardly.

\subsection{Safety and efficiency are separate objectives}

This language stress test shows why the use of a single measure of accuracy is inadequate for clarification systems. The standalone ML system had a much higher accuracy than CLARA and made considerably fewer unnecessary clarifications. However, it did not make any clarifications to answer one question, which was sensitive to its result. There were no critical misses in the case of CLARA but 13 stable queries were unnecessarily interrupted by CLARA. Choosing between these two operating points requires an understanding of the asymmetric costs of the application. For exploratory data analysis, making one extra clarification might be preferable to silently altering the denominator. In the case of a high-throughput user interface, interruptions could decrease usability and lead users to disregard them.

Our findings do not allow us to make an erroneous assertion that execution always performs better than language modeling. In the structured task, both CLARA and ML achieved perfection. In the unstructured stress test, ML performed better.

\subsection{Assay denominator is the main current weakness}

All CLARA mistakes made during the stress test were unnecessary clarifications for the assay-denominator questions. The difficulty with this clarification technique lies in the fact that normal language does not always specify whether the denominator is supposed to include patients who have not been profiled. In this dataset, there were many candidate answers that remained consistent; however, the frozen hybrid strategy did not usually manage to interpret the phrase correctly and clarified conservatively.

A number of improvements can be made without any change to the basic architecture of the system. The parser could be improved by modeling denominator clauses and distinguishing actual omissions from synonyms for profiling. An abstention model could calculate the probability that a translation is in doubt independent of whether or not the executed alternatives were sensitive. The user interface could also adapt to defaults in addition to executing and reporting alternatives. All of these improvements should be tested on freshly written questions because of the contamination of the frozen test by 13 parsing errors.

\subsection{Limitations}

Firstly, CLARA only handles one output type (mutation prevalence) and three ambiguity types. Scientific cancer genomics queries can include other parameters such as copy number alterations, gene expression thresholds, computations, survival rate, drug response, pathways scores, longitudinal samples, and clinical features. 

Secondly, the current research relied on data from eight TCGA PanCancer Atlas datasets in one schema based on the cBioPortal interface. A better validation would include gathering typed specifications from an independent resource like GDC or AACR GENIE and testing the preservation of scientific sense.

Third, the 120-item stress test set was created using LLMs through assignments and manual review by two members of the project. This was done in order to lessen the occurrence of direct template overlap, but this is not the same as prospectively gathering questions from independent cancer researchers. The raters were not able to retain item forms at the item level, so there was no inter-rater reliability measure that can be calculated. 

Fourth, the post-diagnosis templates for verification included the exact 330 executable contrasts and were generated following the corrections of any developmental mistakes. Perfect accuracy indicates the consistency of the corrected pipeline process and not external generalization. 

Additionally, the threshold for the divergence is a practical tolerance and not an absolute biological measure. Finally, CLARA is not a clinical decision-support system. It only analyses query semantics for cohorts without clinical validity or actionability.

\section{CONCLUSION}

Out of 330 comparisons between multiple cohorts, 115 were result sensitive and 215 were stable, and independent realization of the system correctly executed every single candidate interpretation. On the language stress test filtered by hand, all 60 result sensitive cases were identified as such by CLARA while it unnecessarily clarified 13 stable cases. While standalone machine learning was better in general, it made one crucial mistake, showing the trade-off between safety and efficiency rather than superiority of any single approach.

\section*{DATA AND CODE AVAILABILITY}

All code and supplementary materials are available from the corresponding author upon request.

\section*{AUTHOR CONTRIBUTIONS}

\noindent\textbf{Pratyush Shukla:} Conceptualization, Methodology, Software, Validation, Formal Analysis, Investigation, Data Curation, Visualization, Writing: Original Draft, Writing: Review \& Editing.
\textbf{Manveer Singh Tib:} Conceptualization, Methodology, Software, Validation, Formal Analysis, Investigation, Data Curation, Visualization, Writing: Original Draft, Writing: Review \& Editing.
\textbf{Siddhant Garg:} Methodology, Validation, Data Curation, Investigation, Writing: Original Draft, Writing: Review \& Editing.

\section*{ACKNOWLEDGMENTS}

The data presented here is in full or in part based on data provided by the TCGA Research Network (\url{https://www.cancer.gov/tcga}). We acknowledge the contributions made by the TCGA Research Network and the individuals involved in making the PanCancer Atlas possible. Data access and analysis was done using cBioPortal for Cancer Genomics.

\section*{CONFLICT OF INTEREST}

The authors declare no competing interests.

\end{document}